\documentclass{akari2017}

\shorttitle{Cold Diffuse Clouds}
\shortauthors{Gibson et al.}

\begin{document}

\title{Exploring the Dust Population in Cold Diffuse Clouds} 



\correspondingauthor{S. J. Gibson}
\email{steven.gibson@wku.edu}

\author{Steven J. Gibson} 
\affiliation{Department of Physics and Astronomy, Western Kentucky University, 1906 College Heights Blvd., Bowling Green, KY 42101, U.S.A.} 
\affiliation{Academica Sinica Institute of Astronomy and Astrophysics, 
Astronomy-Mathematics Bldg, No.1, Sec. 4, Roosevelt Rd, Taipei 10617, Taiwan, R.O.C.} 
\author{Hiroyuki Hirashita} 
\affiliation{Academica Sinica Institute of Astronomy and Astrophysics, 
Astronomy-Mathematics Bldg, No.1, Sec. 4, Roosevelt Rd, Taipei 10617, Taiwan, R.O.C.} 
\author{Aaron C. Bell} 
\affiliation{Department of Astronomy, Graduate School of Science, The University of Tokyo, Tokyo 113-0033, Japan} 
\author{Mary E. Spraggs} 
\affiliation{Department of Atmospheric and Oceanic Sciences, University of Wisconsin-Madison, 1225 West Dayton Street Madison, WI 53706, U.S.A.} 
\author{Alberto Noriega-Crespo} 
\affiliation{Space Telescope Science Institute, 3700 San Martin Drive, Baltimore, MD 21218, U.S.A.
} 
\author{Sean J. Carey} 
\affiliation{
Infrared Processing and Analysis Center, MC 314-6, California Institute of Technology, 1200 East California Blvd., Pasadena, CA 91125, U.S.A.} 
\author{William T. Reach} 
\affiliation{SOFIA Science Center, MS 232-11, NASA/Ames Research Center, Moffett Field, CA 94035, U.S.A.} 
\author{Christopher M. Brunt} 
\affiliation{School of Physics, University of Exeter, Stocker Road, Exeter, EX4~4QL, United Kingdom} 



\begin{abstract}
The formation and evolution of cold diffuse clouds (CDCs), the parent objects
of dense molecular clouds, affects both the star formation process and that of
larger-scale galactic evolution.  We have begun a pilot study of one CDC's dust
content, with the goal of quantifying the abundances of different types of dust
and relating these to the relative abundance of molecular gas, the cloud's
physical properties, and its general stage of development.  Using photometry
from {\sl AKARI\/} and other surveys, we have extracted a sample spectral
energy distribution (SED) of the CDC dust thermal emission over the
near-infrared to submillimeter range.  The extracted SED closely resembles
others in the literature, confirming our isolation of the cloud emission from
other sources along the sight line.  We plan to fit this SED with dust models
at each position in the cloud, automating our procedure to map out the
structure of this CDC and others.
\end{abstract}



\setcounter{page}{1}



\section{Introduction} 

\begin{figure}[!t]
\begin{center}	
\hspace*{0.7in}
\includegraphics[viewport=0 0 612 512,clip=true,width=3.30in]{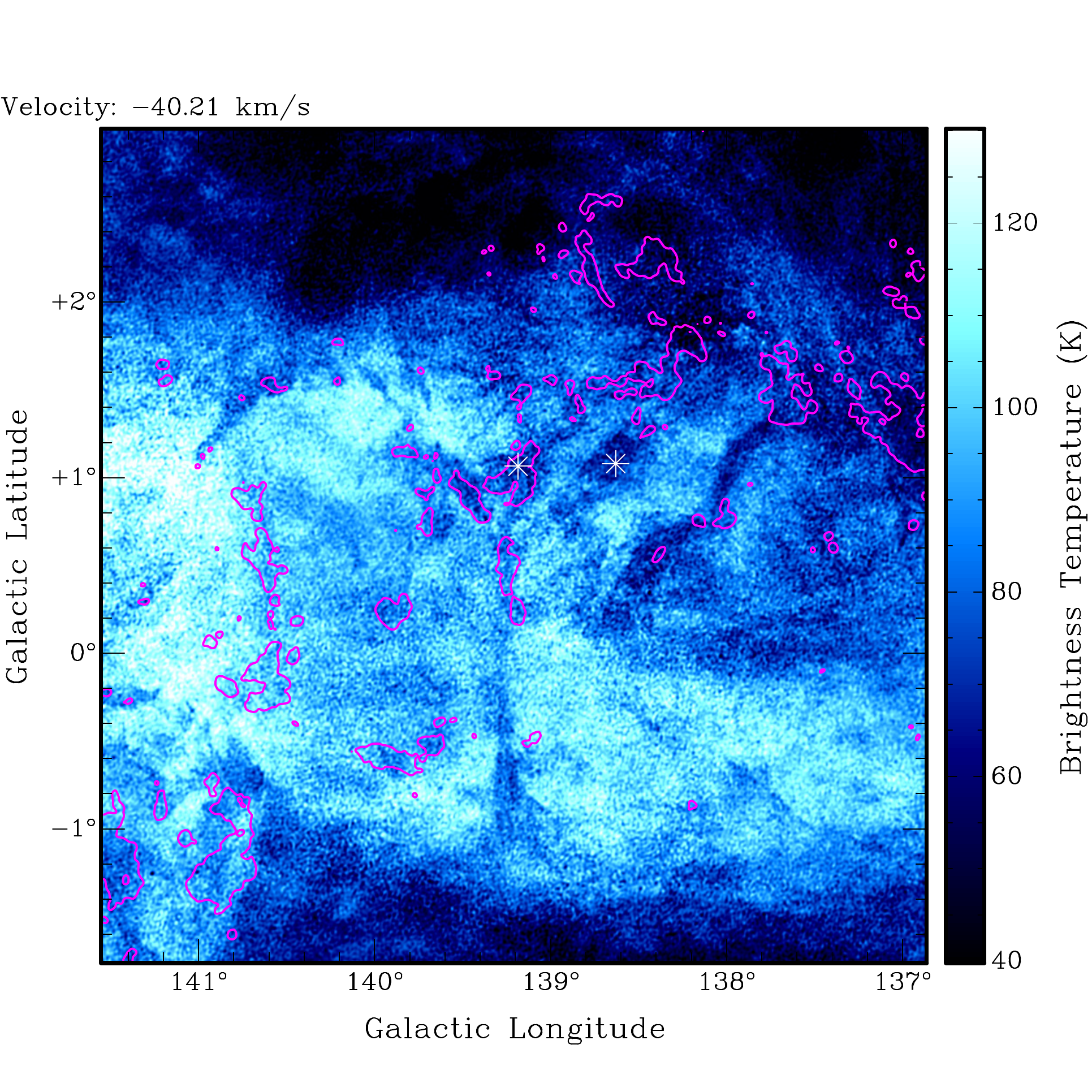}
 \parbox[b]{2.50in}{
  \includegraphics*[viewport=-50 0 560 515,clip=true,width=1.45in]{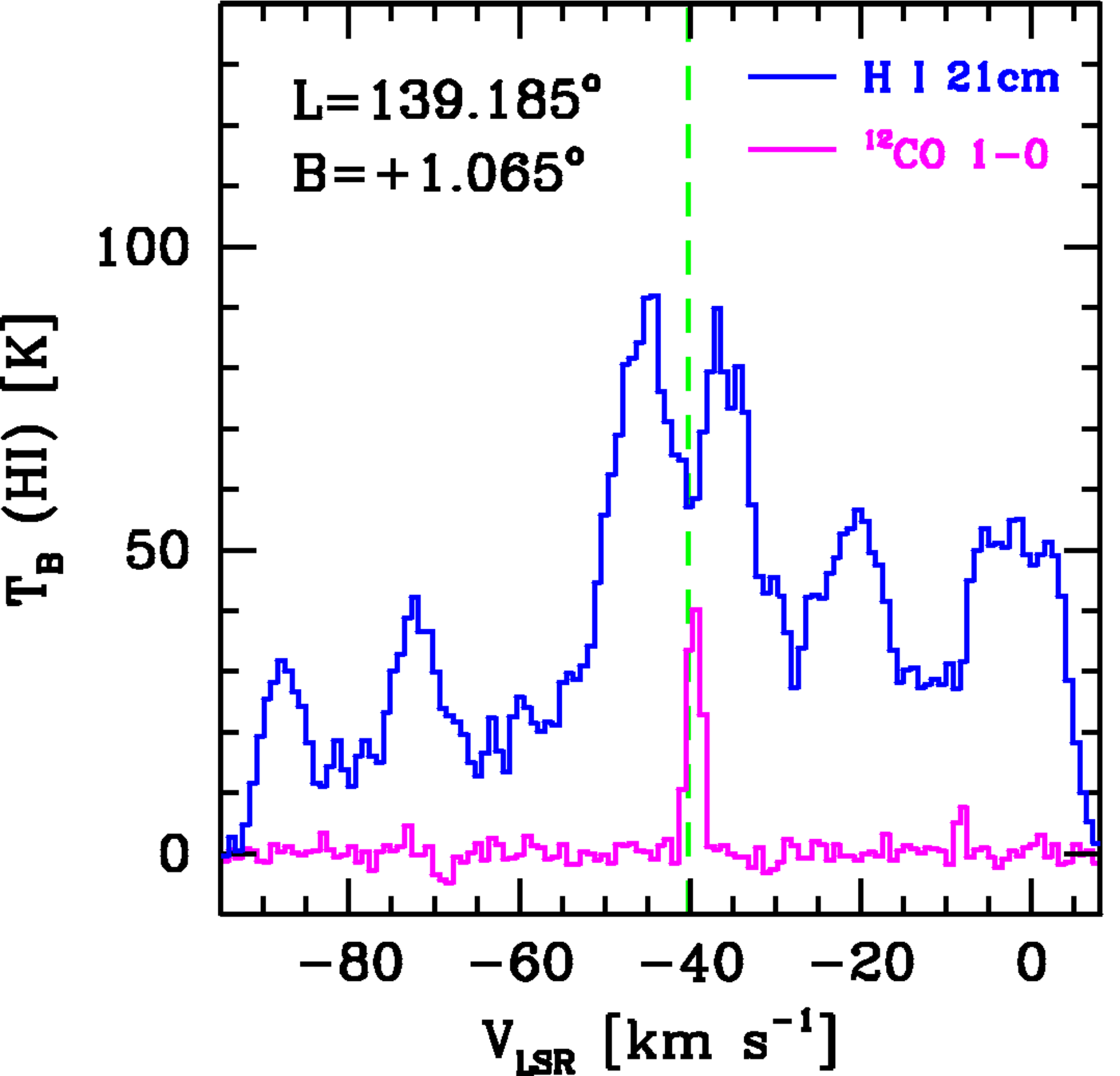}
  \\
  \includegraphics*[viewport=-35 -50 560 515,clip=true,width=1.45in]{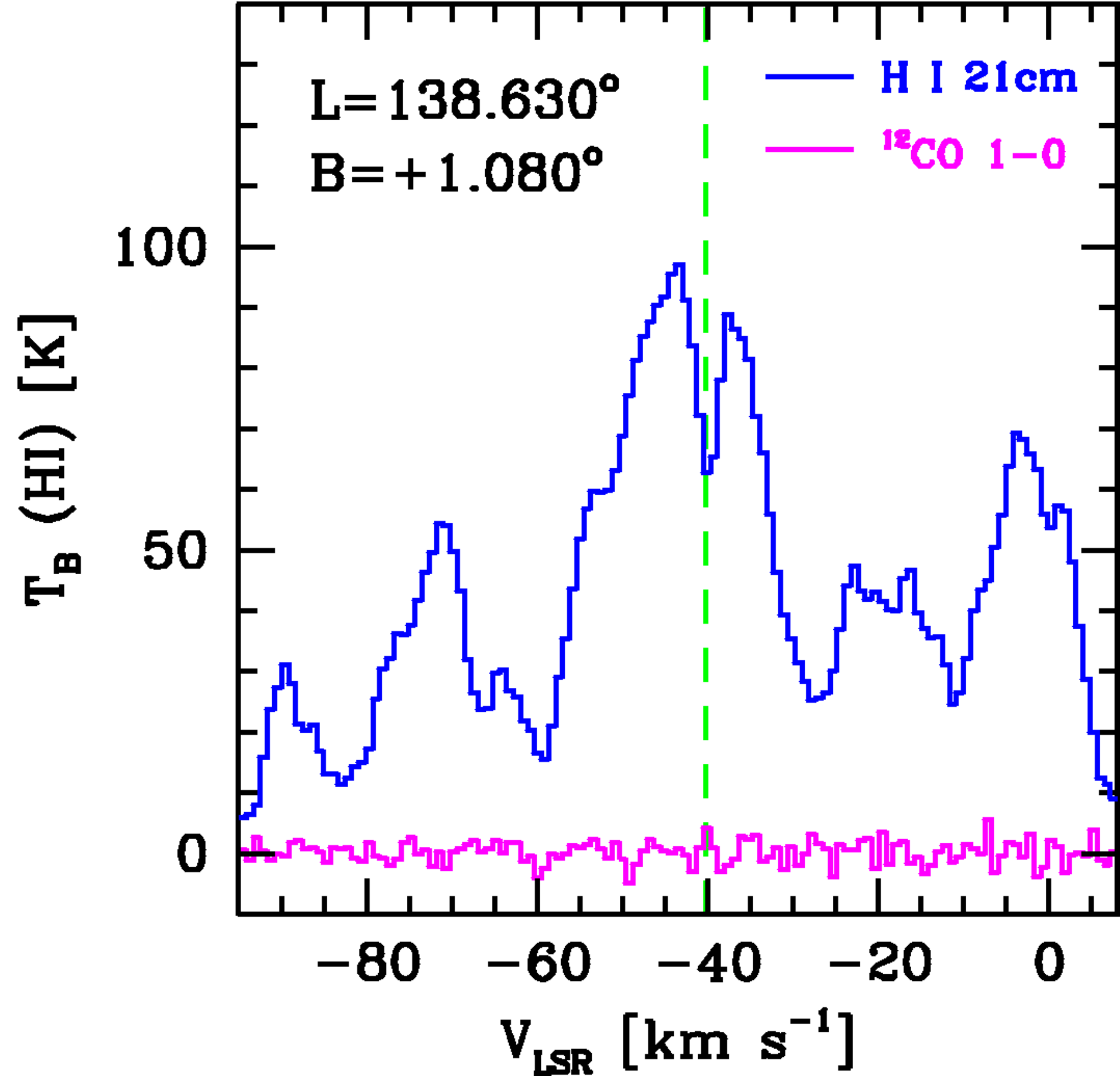}
 }
\caption{
H~{\footnotesize I} self-absorption (HISA) against warmer background
H~{\footnotesize I} emission 
arises from atomic gas that is too cold to explain easily if it is outside
molecular clouds \citep{wolfire_2003}, and yet HISA shadows often appear
separate from CO emission, particularly in the outer Galaxy
(\citealt{gibson_2000}; \citealt{gibson_2010}). The panels above show CGPS
H~{\footnotesize I} emission and absorption (blue; Taylor et al. 2003) and OGS
$^{12}$CO $J=1-0$ emission (magenta; \citealt{heyer_1998}). These clouds are
$\sim 2$~kpc away in the Perseus arm, where they may be forming H$_2$ and CO
downstream of the spiral shock before they become dense enough to form new
stars \citet{gibson_2005}.
}\label{Fig:intro}
\end{center}
\end{figure}

\begin{figure}[!t]
\begin{center}	
\gridline{
  \includegraphics*[viewport=0 5 555 555,clip=true,width=0.3\textwidth]{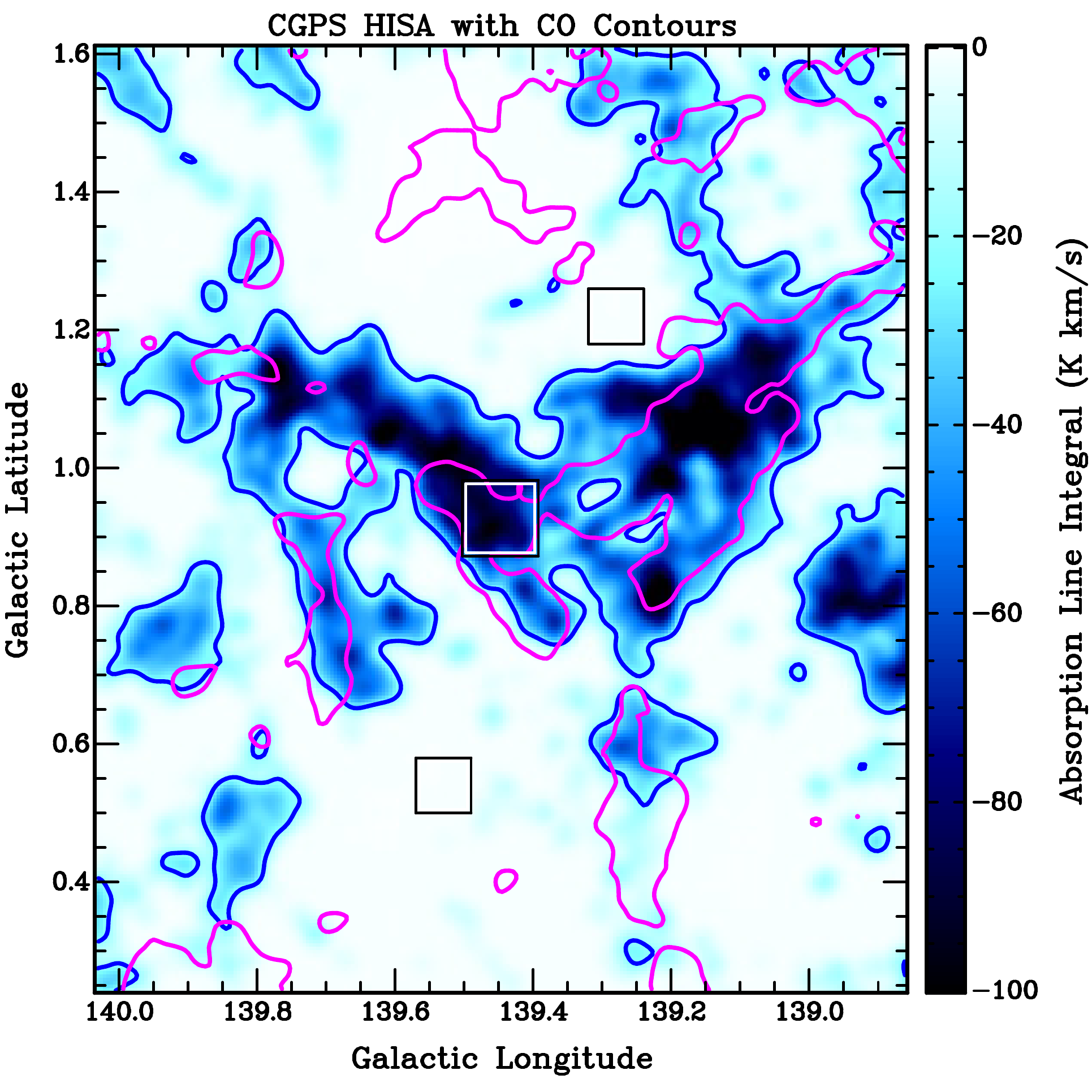}
  \includegraphics*[viewport=0 5 555 555,clip=true,width=0.3\textwidth]{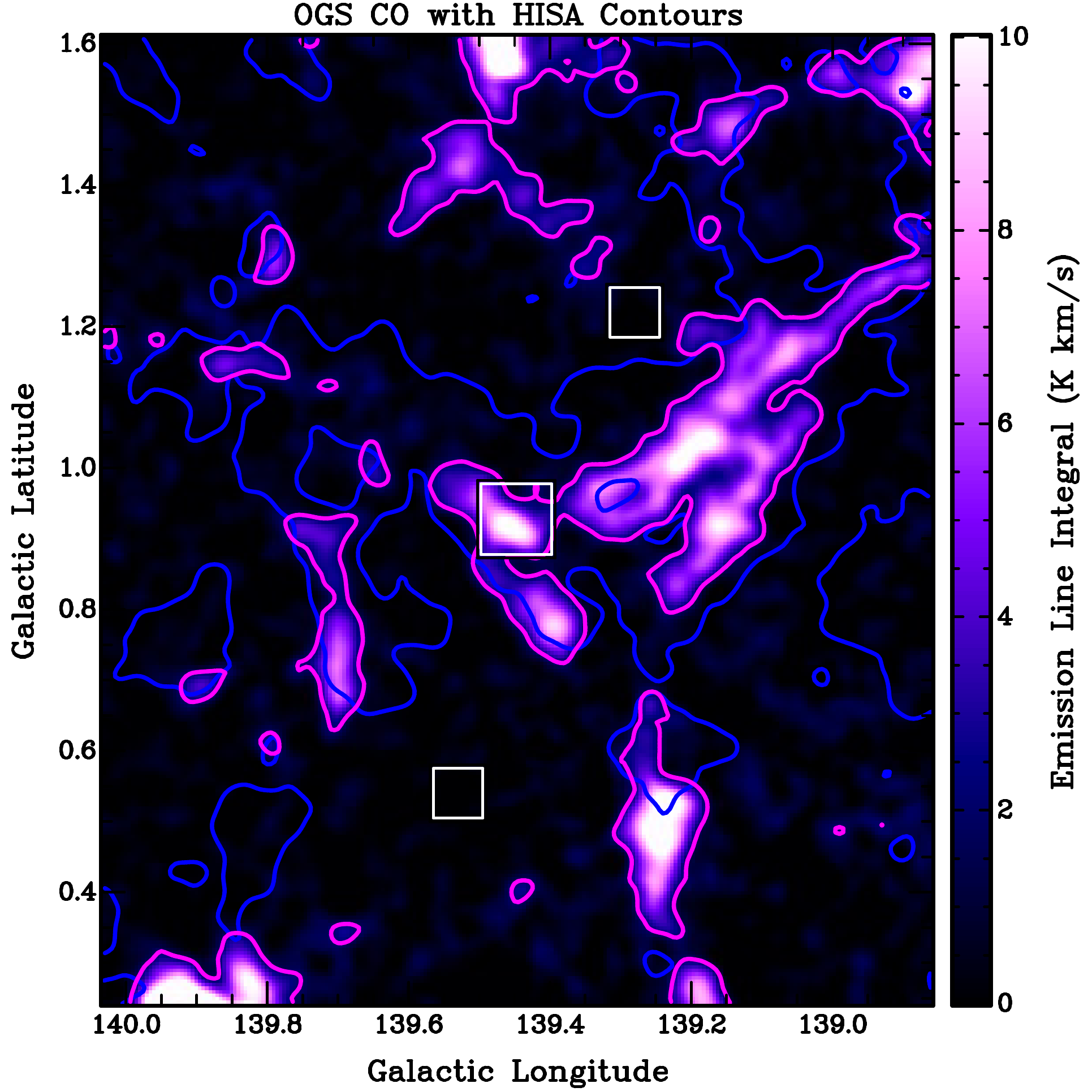}
  \includegraphics*[viewport=0 5 555 555,clip=true,width=0.3\textwidth]{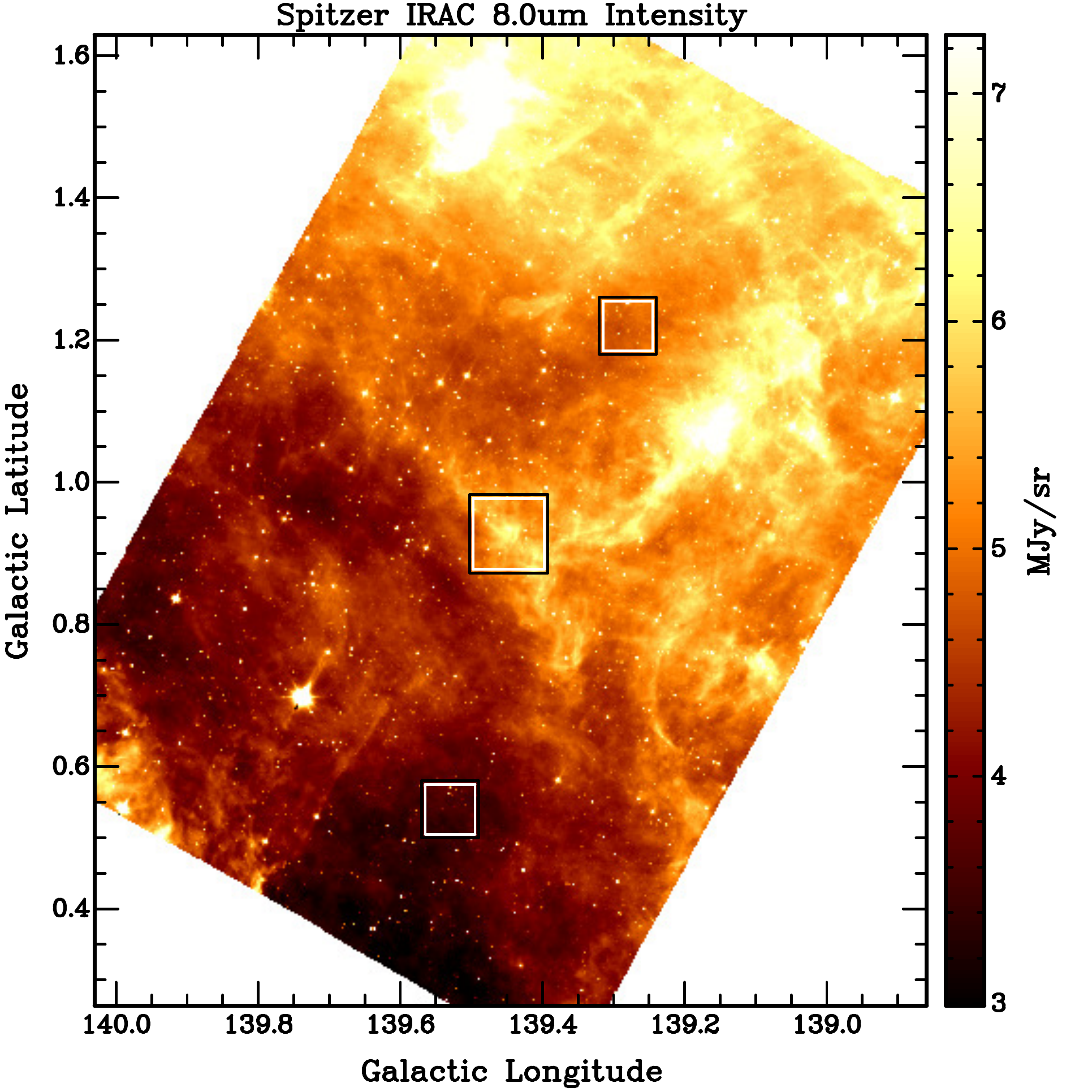}
  }
\gridline{
  \includegraphics*[viewport=0 5 555 555,clip=true,width=0.3\textwidth]{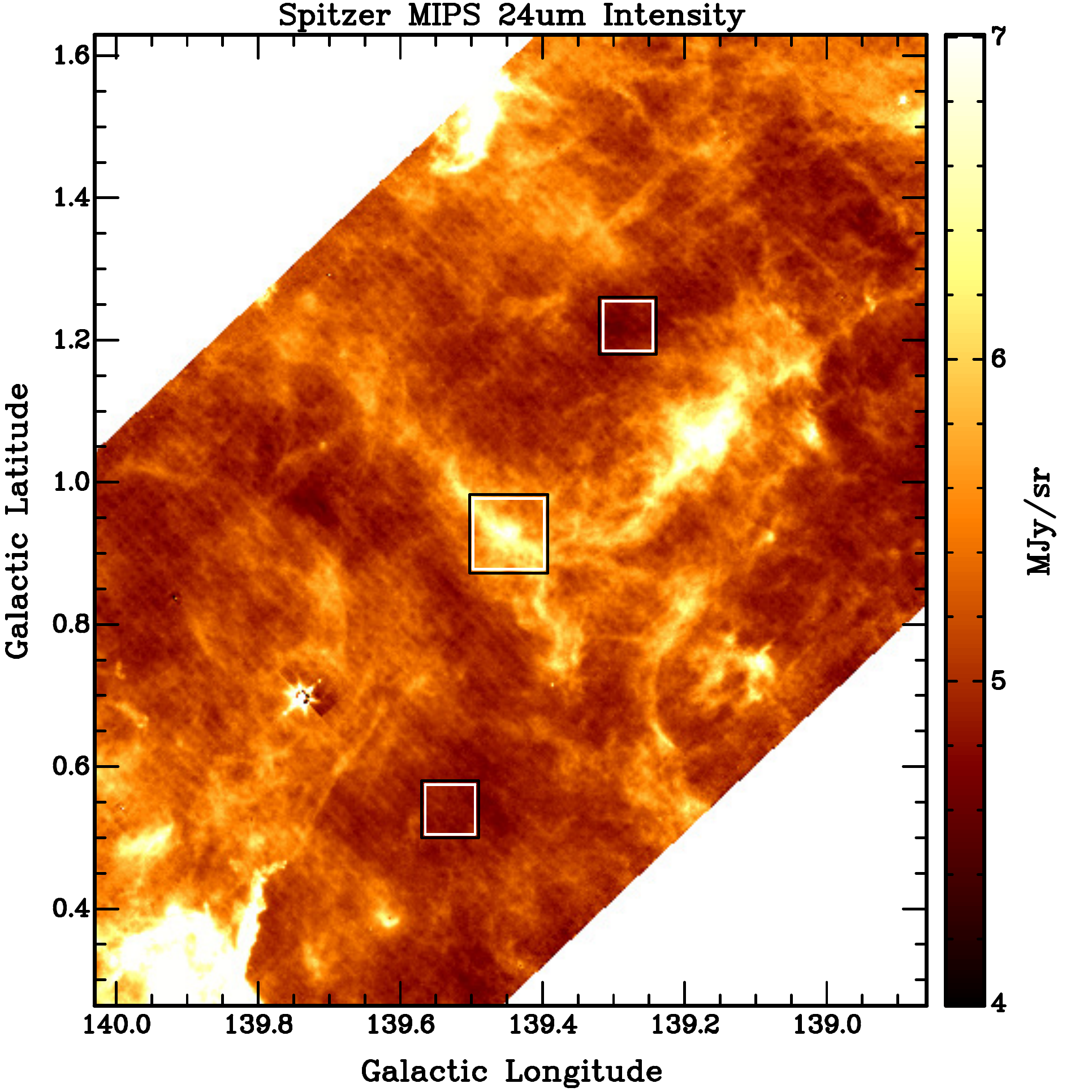}
  \includegraphics*[viewport=0 5 555 555,clip=true,width=0.3\textwidth]{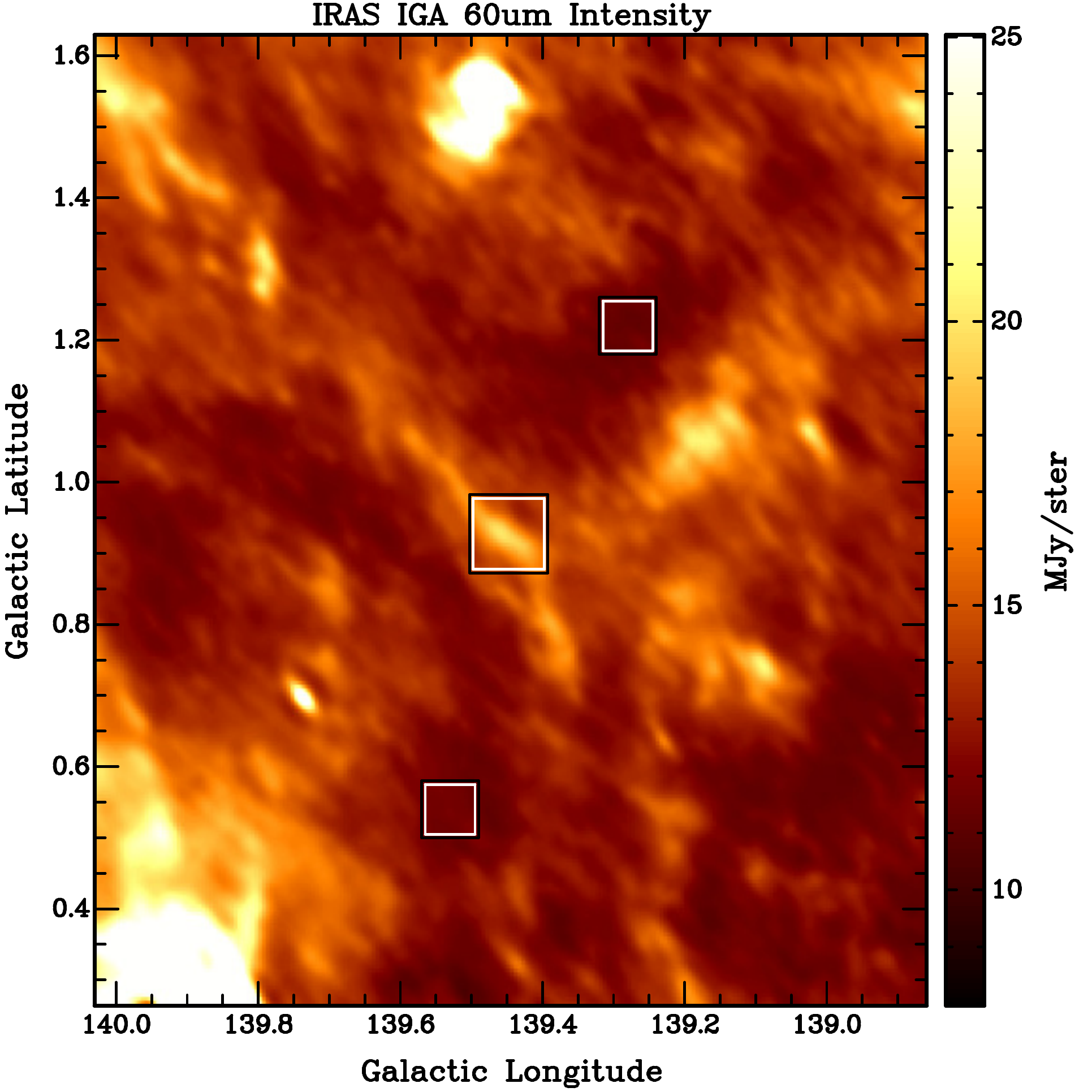}
  \includegraphics*[viewport=0 5 555 555,clip=true,width=0.3\textwidth]{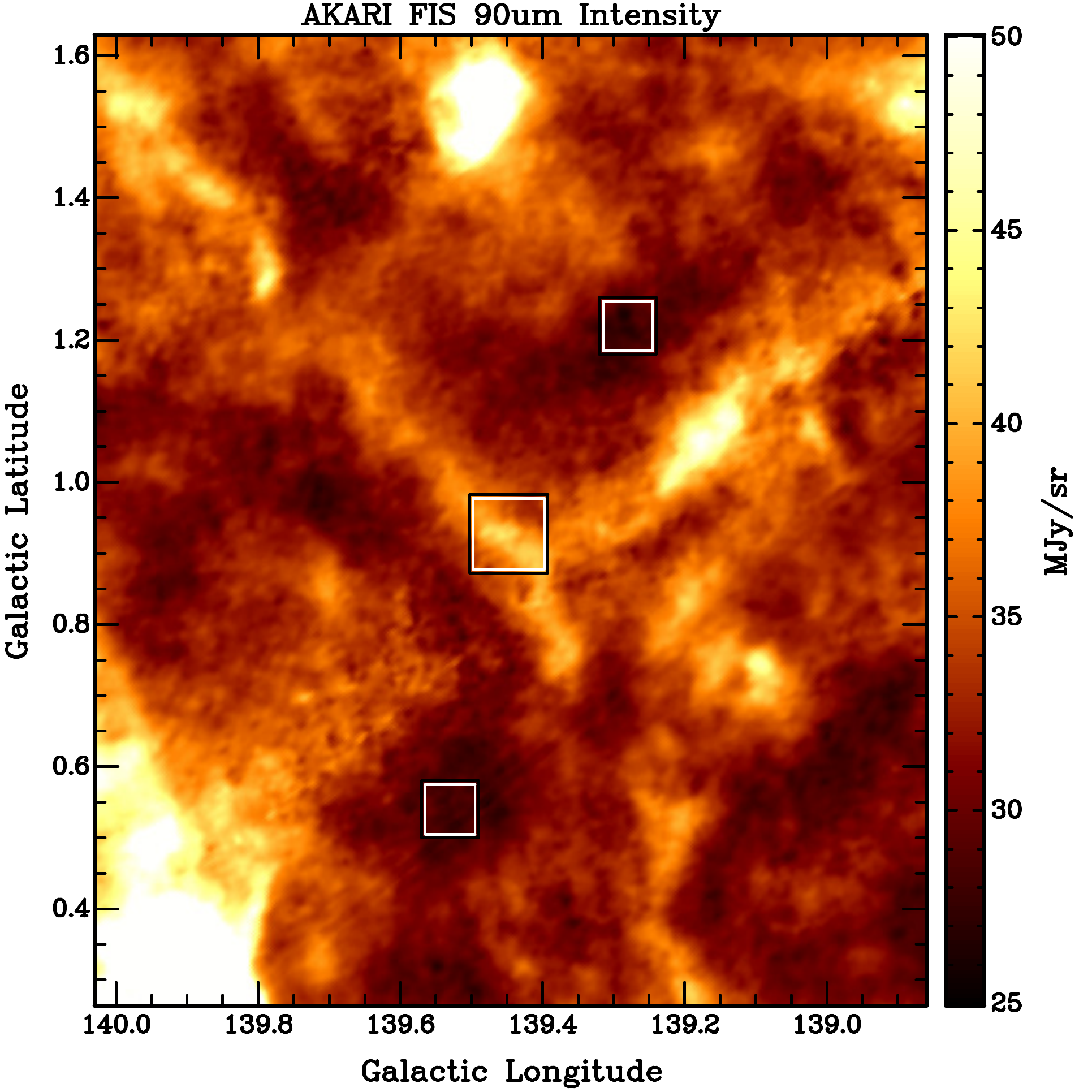}
  }
\gridline{
  \includegraphics*[viewport=0 5 555 555,clip=true,width=0.3\textwidth]{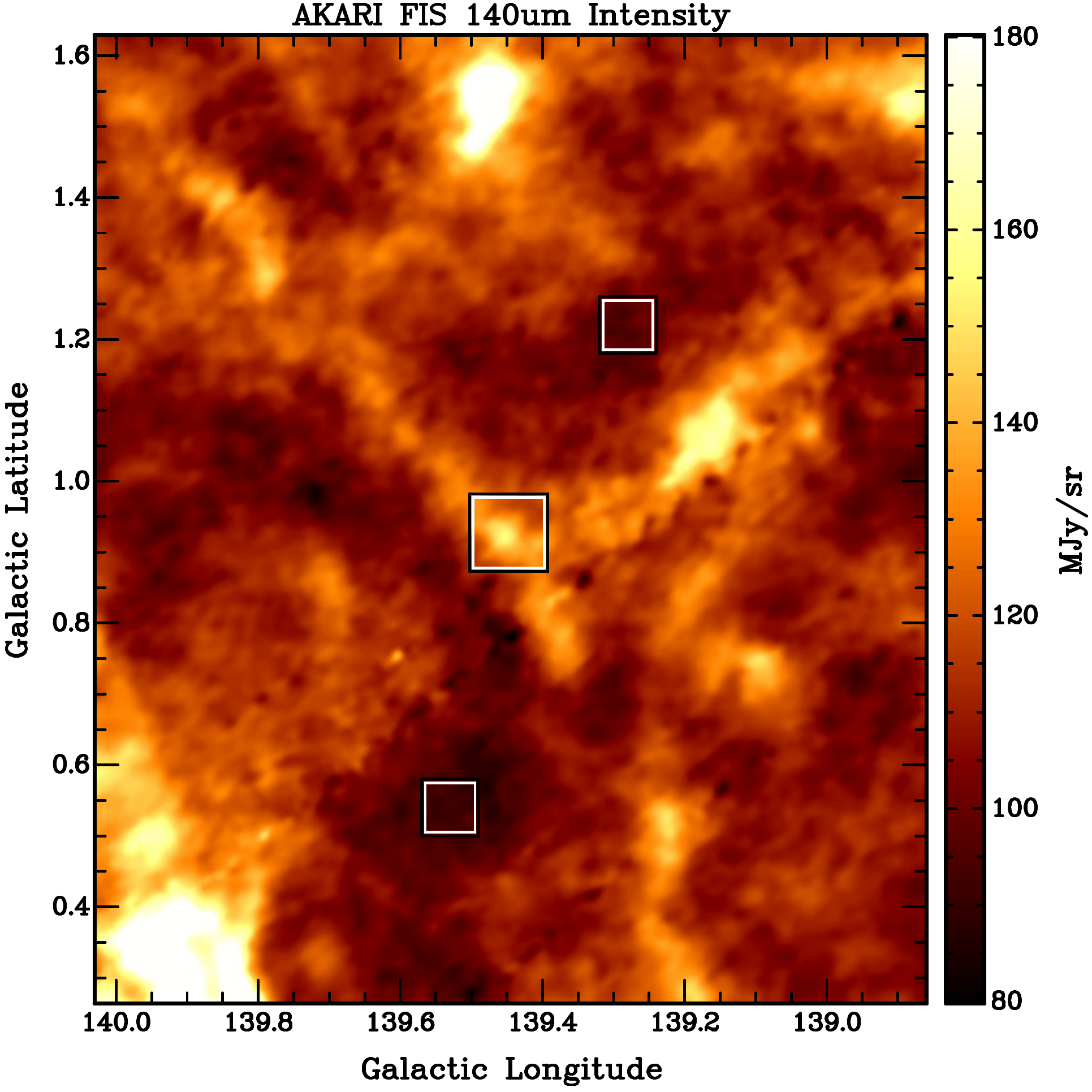}
  \includegraphics*[viewport=0 5 555 555,clip=true,width=0.3\textwidth]{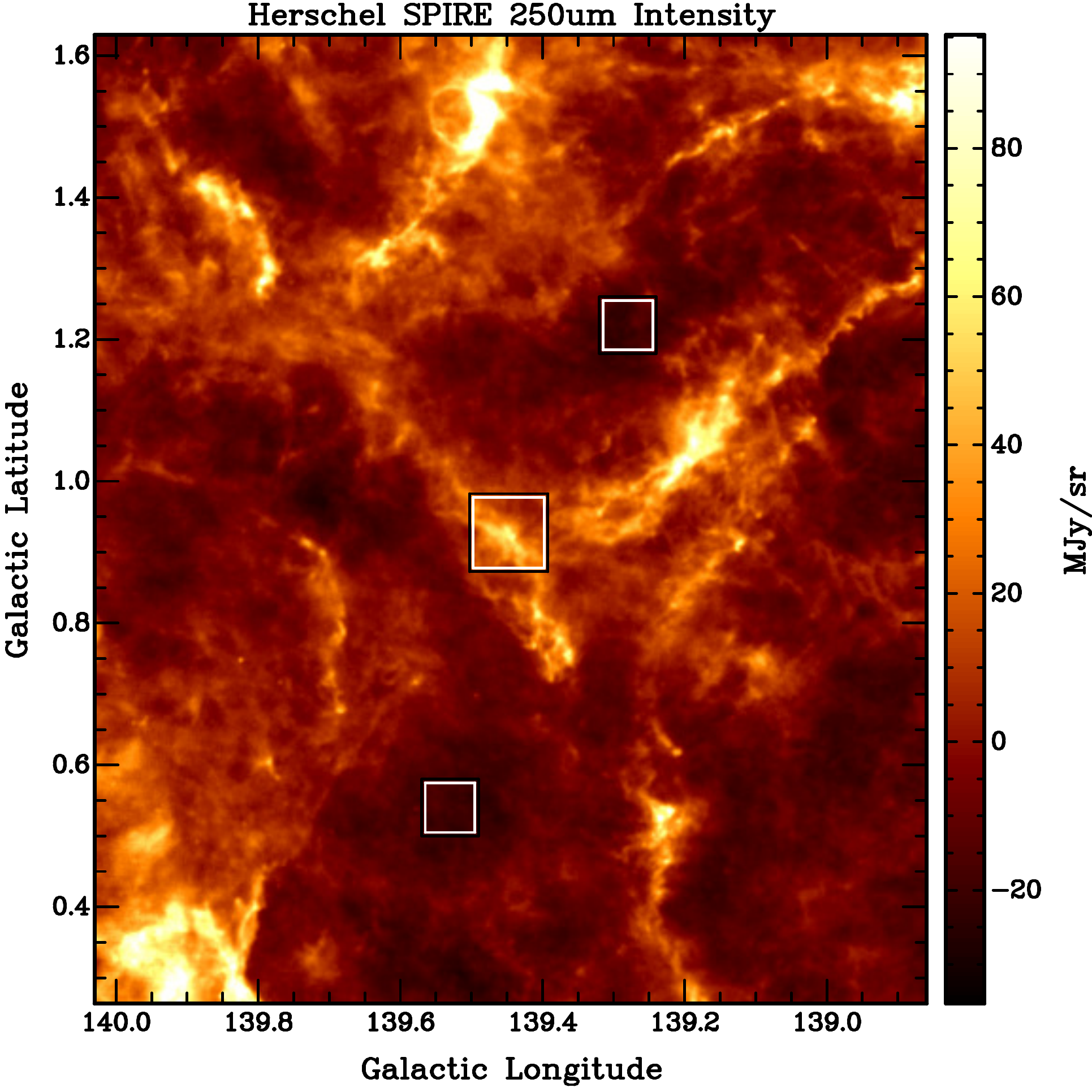}
  \includegraphics*[viewport=0 5 555 555,clip=true,width=0.3\textwidth]{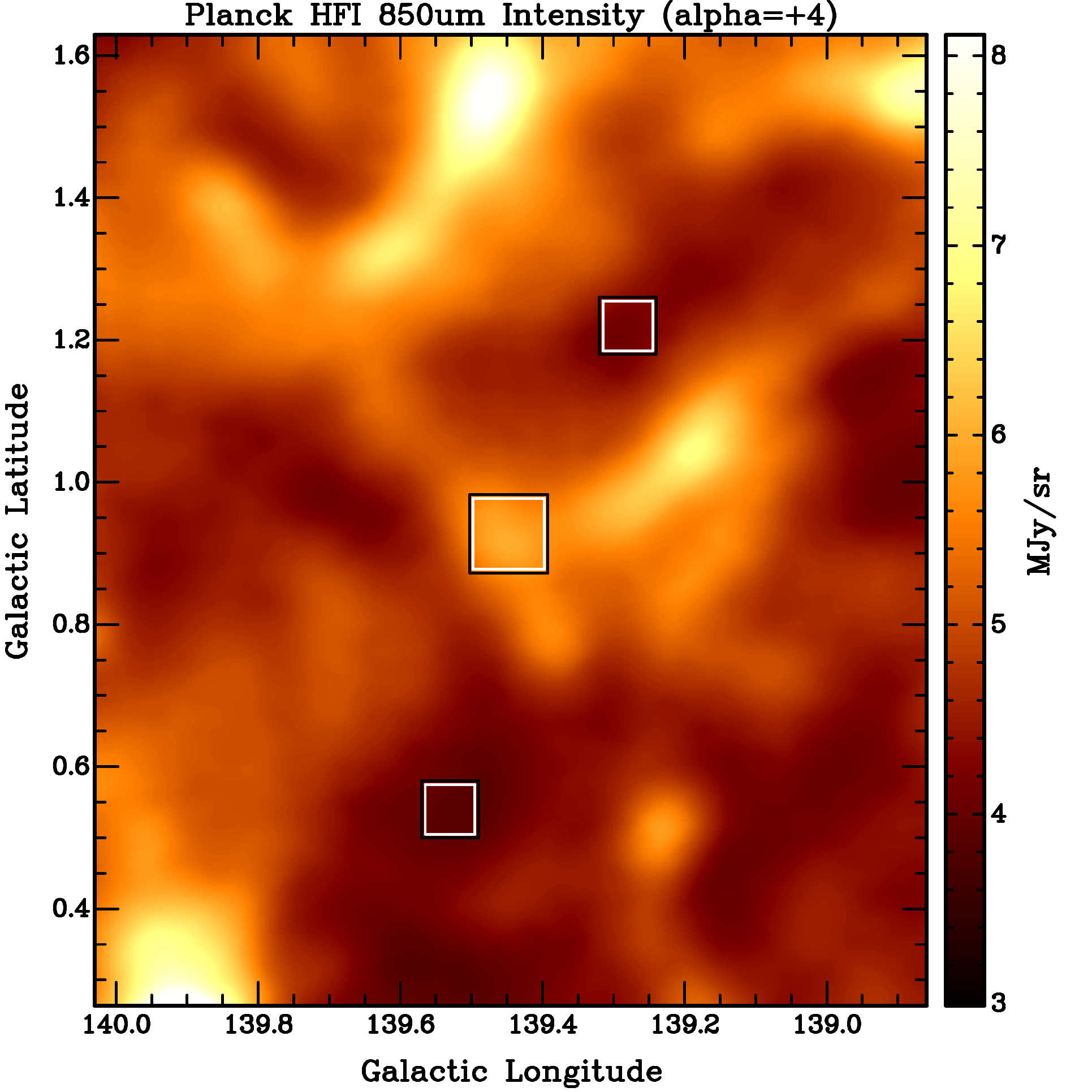}
  }
\caption{
These panels show the region of the current study, which is near the the center
of {\bf Figure~\ref{Fig:intro}}. {\it Top left:\/} smoothed line integral of
extracted HISA. {\it Top center:\/} line integral of associated CO emission.
Other panels show a subset of the dust image data assembled for the same region
for $\lambda = 8$, 24, 60, 90, 140, 250, \& $850 {\rm \, \mu}$m; a full list is
given in {\bf Table~\ref{Tab:surveys}}. The boxes mark one ON- and two
OFF-cloud areas from which dust emission photometry data were extracted with a
$3\sigma$-clipped median statistic; the cloud emission was isolated as the ON
minus the averaged OFF brightness, with RMS scatter used as a proxy for
measurement uncertainty.
}\label{Fig:maps}
%
\end{center}
\end{figure}

\begin{figure}[!ht]
\begin{center}	
\includegraphics[width=3.0in]{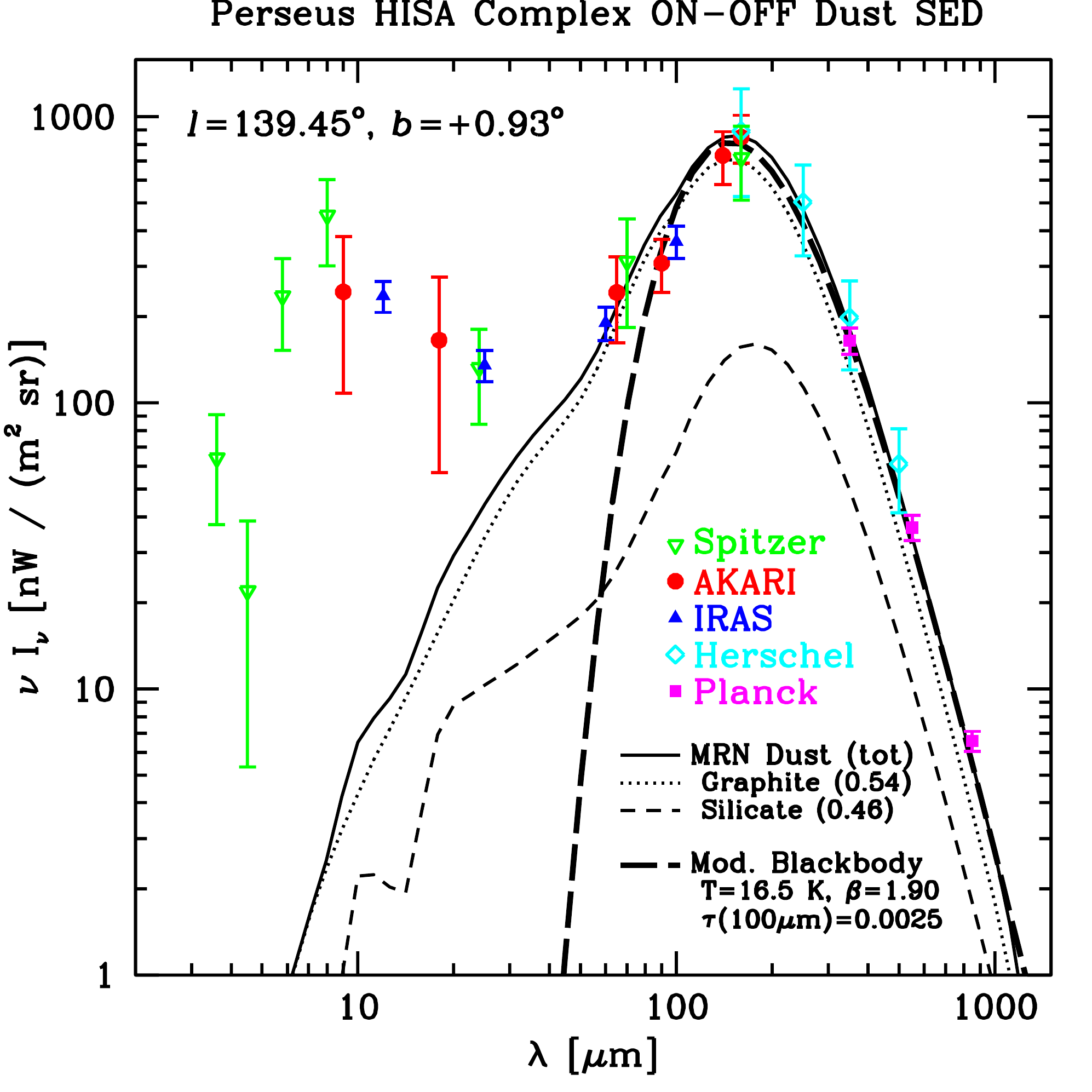}	
\caption{
Extracted ON-OFF photometry for the sample sight line taken at the positions
marked in {\bf Figure~\ref{Fig:maps}} from each of the data sets listed in {\bf
  Table~\ref{Tab:surveys}}. A simple SED model with MRN \citep{mathis_1977}
grains is shown for a standard interstellar radiation field
\citep{mathis_1983}, gas/dust ratio, and metallicity. This model does not have
enough VSGs and lacks PAHs, but it fits the longer-wavelength data better than
a modified blackbody.  Visual comparison to various other SED measurements and
models in the literature (e.g., \citealt[Fig.~2]{compiegne_2011})
highlights the clear detection of many shorter-wavelength features.
%
}\label{Fig:sed}
\end{center}
\end{figure}

Cold diffuse clouds (CDCs) are a key transitional phase between the warm,
tenuous, ambient interstellar medium of mostly-neutral atomic gas and the
colder, denser molecular clouds needed for star formation. However,
understanding exactly how CDCs evolve is hampered by observational challenges,
as much of their gas is "dark" in H~{\footnotesize I} 21~cm and CO 2.6~mm
spectral line emission surveys, while their dust thermal continuum emission can
be faint and hard to distinguish from denser clouds in the same sight line. We
have developed methods to identify CDCs using
H~{\footnotesize I} self-absorption from their cold atomic gas (HISA; {\bf
  Figure~\ref{Fig:intro}}; \citealt{gibson_2010}), and to isolate their dust
emission from confusing backgrounds \citep{spraggs_2016} in order to measure
their spectral energy distributions (SEDs) in infrared and sub-millimeter
surveys like {\sl AKARI\/}, {\sl IRAS\/}, and {\sl Planck\/} (see {\bf
  Table~\ref{Tab:surveys}}). Using SED model fits, we aim to constrain the dust
temperature, size distribution, composition, and column density and then
examine how these may relate to the H~{\footnotesize I} and H$_2$ content of
the CDCs. For example, emission from large grains in thermal equilibrium 
relates to both the gas total column density and shielding of the cloud
interior, while very small grain (VSG) surfaces are important sites for H$_2$
formation and photoelectric heating. SED studies can thus inform us of the
evolution of the grains within the clouds as well as their effect on the
clouds' physical state. We have assembled a number of IR/submm maps of a sample
CDC in the outer Galaxy ({\bf Figure~\ref{Fig:maps}}) and have begun
investigating simple SED model fits for one sight line in this object ({\bf
  Figure~\ref{Fig:sed}}).

\renewcommand{\arraystretch}{0.9}
\begin{table}[!b]
\centering
\caption{Data Sets Used in This Study}
\begin{tabular}{|c|c|c|c|l|}
 \hline
Observatory & Instrument & Spectral Band & Beam Size & Surveys/References \\
 \hline\hline
DRAO & ST+26m & H~{\footnotesize I} 21~cm line & 60$''$ & CGPS$^{1}$ \\
FCRAO & 14m & $^{12}$CO 2.6~mm line & 45$''$ & OGS$^{2}$, EOGS$^{3}$ \\
 \hline
{\sl Spitzer\/} & IRAC & 3.6, 4.5, 5.8, $8.0\,\mu$m & 2$''$ & (this work)$^{4}$ \\
{\sl AKARI\/} & IRC & 9, $18\,\mu$m & 6$''$ & {\sl AKARI\/} Mid-IR All-Sky$^{5}$ \\
{\sl IRAS\/} & SA & 12, 25, 60, $100\,\mu$m & $70 - 260''$ & IGA$^{6}$, MIGA$^{7}$, IRIS$^{8}$ \\
{\sl Spitzer\/} & MIPS & 24, 70, $160\,\mu$m & $7 - 47''$ & (this work)$^{4}$ \\
{\sl AKARI\/} & FIS & 65, 90, 140, $160\,\mu$m & $63 - 88''$ & {\sl AKARI\/} Far-IR All-Sky$^{5}$ \\
{\sl Herschel\/} & PACS & 70, $160\,\mu$m & $6 - 13''$ & Hi-GAL$^{10}$ \\
{\sl Herschel\/} & SPIRE & 250, 350, $500\,\mu$m & $18 - 36''$ & Hi-GAL$^{10}$ \\
{\sl Planck\/} & HFI & 350, 550, $850\,\mu$m & 300$''$ & {\sl Planck\/} Data Release 2$^{11}$ \\
 \hline
    \multicolumn{5}{l}{\footnotesize 
	$^{1}$CGPS: \citet{taylor_2003};
	$^{2}$OGS: \citet{heyer_1998};
	$^{3}$EOGS: \citet{mottram_2010}
	} \\
    \multicolumn{5}{l}{\footnotesize 
	$^{4}$\citet{bell_2012};
	$^{5}$IRC: \citet{ishihara_2010};
	$^{6}$FIS: \citet{doi_2015};
	$^{7}$IGA: \citet{cao_1997}
	} \\
    \multicolumn{5}{l}{\footnotesize 
	$^{8}$MIGA: \citet{kerton_2000};
	$^{9}$IRIS: \citet{mamd_2005}
	} \\
    \multicolumn{5}{l}{\footnotesize 
	$^{10}$Hi-GAL: \citet{molinari_2010};
	$^{11}${\sl Planck\/} DR2: \citet{planck_2016_1,planck_2016_10}
	} \\
\end{tabular}
\label{Tab:surveys}
\end{table}

\section{Current Results}

Careful differencing of infrared brightnesses in ON-cloud and OFF-cloud
positions (marked in {\bf Figure~\ref{Fig:maps}}) yields an SED curve that
looks reasonably similar to others in the literature, which should allow
detailed constraints on dust SEDs vs.\ position within CDCs, even for sight
lines that include other emission sources.

A preliminary MRN \citep{mathis_1977} model SED fit of these data ({\bf
  Figure~\ref{Fig:sed}}) shows that the long-wavelength ($\lambda > 80 \,
\mu$m) brightness is consistent with isothermal large-grain emission in a
modified blackbody.  At wavelengths just below this ($\lambda \sim 60-70 \,
\mu$m), very small grains can be fit with a simple MRN power-law distribution.
But at shorter wavelengths ($\lambda < 50 \, \mu$m), the MRN model lacks
sufficient VSG emission and has no polycyclic aromatic hydrocarbons (PAHs) or
similar components.  A more sophisticated model is obviously needed.

The MRN model H-atom column through the cloud ($N_{\rm H} = N_{\rm HI} + 2
N_{\rm H_2} = 1.3 \times 10^{21} {\rm \, cm^{-2}}$) is broadly consistent with
HISA radiative transfer constraints \citep{gibson_2000} but below that found in
combined HISA+CO spectral line analyses (e.g., $N_{\rm H} > 2.4 \times 10^{21}
{\rm \, cm^{-2}}$; \citealt{klaassen_2005}).  However, half of either column
may still allow sufficient shielding of the cloud core to reduce heating by
external radiation ($A_V \sim 0.3 - 0.6$~mag for $R_V \sim 3$ with the
\citealt{bohlin_1978} conversion factor), thus ``hiding'' some dust from an
isothermal SED fit.  This possibility is under investigation.

\section{Future Work}

Several steps are planned to put the analysis on a more solid footing.  In
addition to incorporating other available short-wavelength data (e.g., {\sl
  WISE\/}; \citealt{wright_2010}) and comparing fits for models with more
VSG/PAH content (e.g.,
\citealt{draine_2007}; \citealt{galliano_2011}; \citealt{jones_2017}), we will
measure the RMS noise more rigorously, convolve all maps to a common angular
resolution, and fit the observed data with synthetic photometry of our SEDs
using instrument spectral response curves.  We will also automate the OFF
region selection and SED fitting map results vs.\ position throughout a given
cloud, using the OFF position selection and interpolation algorithm developed
by \citet{spraggs_2016}, so that gas and dust properties can be studied
spatially.

Beyond the basic single-zone dust model, we will explore CDC core shielding
effects resulting from different cloud $N_{\rm H}$ columns to check the
robustness of the column fits.  We will also consider possible dust evolution
and its relation, if any, to the CDC evolution or the larger environment.  The
last will be aided by similar analyses of other HISA CDCs in different parts of
the Galaxy, most of which have already been identified \citep{gibson_2015}.
Finally, through the use of available all-sky survey data sets, off-plane CDCs
showing narrow-line H~{\footnotesize I} emission rather than HISA can be
similarly analyzed to provide a more complete view of the CDC population.

\subsection*{Acknowledgments}
This work has been supported by the U.S. National
Science Foundation, NASA, Western Kentucky University, and the Academica Sinica
Institute of Astronomy and Astrophysics.
This research makes use of observations with {\sl AKARI\/}, a JAXA project with
the participation of ESA.  Many data sets were obtained through the NASA/IPAC
Infrared Science Archive (IRSA).  Data analysis and presentation were aided by
the
Montage, Miriad, SuperMongo, and Karma software packges.




\end{document}